\documentclass[flushrt]{emulateapj}

\begin{document}

\title{Two Lensed $\lowercase{z} \simeq 3$ Lyman Break Galaxies Discovered in the SDSS Giant Arcs Survey}
\author{Benjamin P. Koester\altaffilmark{1,2}, Michael D. Gladders\altaffilmark{1,2,3}, Joseph F. Hennawi\altaffilmark{4}, 
Keren Sharon\altaffilmark{1,2}, Eva Wuyts\altaffilmark{1}, J.R. Rigby\altaffilmark{5}, Matthew B. Bayliss\altaffilmark{1,2}, 
Hakon Dahle\altaffilmark{7}}
\altaffiltext{1}{Department of Astronomy and Astrophysics, University of Chicago, Chicago IL 60637, USA}
\altaffiltext{2}{Kavli Institute for Cosmological Physics, The University of Chicago, Chicago IL 60637, USA}
\altaffiltext{3}{Visiting Associate, Observatories of the Carnegie Institution of Washington, Pasadena, CA 91101}
\altaffiltext{4}{Max Planck Institut fur Astronomie, Konigstuhl 17, D-69117 Heidelberg, Germany}
\altaffiltext{5}{Observatories, Carnegie Institution of Washington, 813 Santa Barbara St., Pasadena, CA 91101, USA ; Spitzer Fellow}
\altaffiltext{6}{Institute of Theoretical Astrophysics, University of Oslo, P.O. Box 1029, Blindern, N-0315 Oslo, Norway}

\begin{abstract}
We report the discovery of two strongly-lensed $z\sim 3$ Lyman Break 
Galaxies (LBGs) discovered as $u$-band dropouts as part of the SDSS 
Giant Arcs Survey (SGAS). The first,
SGAS J122651.3+215220 at $z=2.9233$ is lensed by one of several sub-clusters,
SDSS J1226+2152, in a complex massive cluster at $z=0.43$. 
Its ($g,r,i$) magnitudes are ($21.14,20.60,20.51$)
which translate to surface brightnesses, $\mu_{g,r,i}$, of ($23.78,23.11,22.81$). 
The second, SGAS J152745.1+065219, is an LBG at $z=2.7593$ lensed by the foreground 
SDSS J1527+0652 at $z=0.39$, with $(g,r,z)$=($20.90,20.52,20.58$) and 
$\mu_{g,r,z}$=($25.15,24.52,24.12$). Moderate resolution spectroscopy confirms 
the redshifts suggested by photometric breaks, and shows both absorption and 
emission features typical of LBGs. Lens mass models derived from 
combined imaging and spectroscopy reveal that SGAS J122651.3+215220 is a highly 
magnified source ($M \simeq 40$), while SGAS J152745.1+065219 is magnified by
no more than $M\simeq 15$. Compared to LBG survey results \citep{steidel03},
the luminosities and lensing-corrected magnitudes
suggest that SGAS J122651.3+215220 is among the faintest $\simeq 20\%$ of 
LBGs in that sample. SGAS J152745.1+065219, on the other hand, appears to be more representative
of the average LBG, similar to the ``Cosmic Eye''.

\subjectheadings{galaxies: formation, galaxies: high-redshift, gravitational lensing, cosmology: early universe}

\end{abstract}

\section{Introduction}
In the sequence of gravitational collapse, heating,
stellar ignition and death that govern the evolution of 
baryons in the earliest overdensities, Lyman Break Galaxies (LBGs) serve as
high-redshift way points on the path to the $z \sim 0$ galaxies
we observe today \citep[e.g.,][]{adelberger98,steidel03}. 
In star-forming galaxies at $z \sim 3$, the Lyman continuum 
break at 912~\AA~ resides in blue optical bands ($u,g$), while the 
continuum itself can be detected at redder wavelengths ($g,r$).
These features have motivated the construction
of photometric surveys that rely on this ``dropout'' technique
\citep[][and references therein]{steidel95} 
to select hundreds of likely LBGs at $z \sim 3$ \citep{steidel96a,adelberger03,steidel03}.  
However, typical samples of LBGs consist
of tens of objects with fluxes too faint to permit detailed 
spectroscopic observations \citep[e.g.,][]{nesvadba06} of individual
systems. \citet{shapley03} addressed this shortcoming by creating
a high S/N composite spectra from low S/N spectra of $\simeq 1000$ LBGs
to infer the properties of the average LBG. 

Alternatively, since the discovery of MS1512-cB58 at $z=2.7$ 
\citep[cB58;][]{yee96}, 
strongly-lensed LBGs magnified tens of times
have offered a high S/N window \citep{teplitz00,pettini02} into the conditions
of individual star-forming galaxies when the Universe was less than
2 Gyrs old. For example, the presence of ultraviolet absorption lines
in cB58 \citep{pettini00,pettini02} reveal a chemically diverse
ISM, which suggests that most of the metal enrichment occurred within the 
previous $\sim 300$ Myr, and that the energetic star-formation
drives a bulk outflow of the ISM that exceeds the star-formation
rate. Studies of cB58 with \emph{Spitzer} \citep[e.g][]{siana08} 
have recently brought the IR observations into the picture, 
suggesting that the UV-inferred star-formation rate is a factor 
of $\sim 3-5$ lower than that measured in the IR. 

Following the discovery of cB58, three more 
strongly-lensed LBGs have been added: 1E0657-56 at $z=3.24$ \citep{mehlert01}, 
the 8 o'clock arc \citep{allam07} at $z=2.73$, and the 
Cosmic Eye at $z=3.07$ \citep{smail07}. Further 
spectroscopy \citep{finkelstein09}, space-based IR 
\citep{siana09}, and millimeter \citep{coppin07} observations
of the latter two systems have begun to fill out our
understanding of $z=3$ LBGs but still leave many 
unanswered questions.
   
In this Letter, we report the discovery of two more strongly-lensed
LBGs, and include a description of their basic properties. Given the burgeoning 
rate of discovery of large samples of lensed sources \citep[e.g.,][]{gladders05,cabanac07,hennawi08} we 
refrain from assigning nicknames to the sources discussed here; instead, we introduce designations 
of the form SGAS JXXXXXX+XXXXXX to denote lensed sources and SDSS J????+???? to denote the 
cluster lenses. Where necessary, we assume a flat
$(\Omega_m,\Omega_\Lambda)=(0.3,0.7)$ cosmology with 
$H_0=70$ km s$^{-1}$ Mpc$^{-1}$.
\newpage
\section{Data}
\subsection{SGAS}
Beginning in May 2005, we initiated the SDSS Giant Arcs Survey
\citep[SGAS;][]{hennawi08}, a blind survey for
strong-lensing systems in massive
clusters at $0.1 < z < 0.6$ detected in the 
Sloan Digital Sky Survey \citep[SDSS;][]{york00} using 
the SDSS adaptation of the Red-Sequence Cluster algorithm 
\citep{gladders00}. Clusters
are blindly chosen and imaged in the $g$-band at 
2m to 4m-class telescopes for 600s in $<1''$ seeing and then visually inspected
for giant arcs. For the most promising sources, follow-up imaging and 
low resolution spectroscopy at 8m-class telescopes is used to study the sources
in detail, and to secure the redshift of the putative lensed source. 
In some cases we have obtained medium resolution spectroscopy
to begin to explore the intrinsic properties of the arcs themselves. 
As of May 2009, the blind survey
includes nearly 600 clusters with initial imaging follow-up, 
tens of which are new strong lenses. 

SGAS 152745+065219 was discovered in May 2005 on the 3.5m 
WIYN telescope at Kitt Peak, and presented as a probable
lensing system in \citet{hennawi08}. As part of the 
spectroscopic follow-up program (see below) we obtained
redshifts for 13 cluster members that set the redshift
of the cluster lens at $z=0.39$, 
with $\sigma_v \simeq 908$ km s$^{-1}$. In December 2007, 
SGAS 122651+215220 was discovered at the 2.5m Nordic
Optical Telescope. The lens photometric redshift, $z=0.43$,
for SDSS J1226+2152 has since been confirmed by 
several objects from the SDSS spectroscopic survey and 
GMOS spectroscopy. Notably, a separate arc in this same system
was deemed a ``probable'' lensed galaxy by 
\citet{wen09} (NSCS J122648.0+215157).  
The lens  itself is one of several sub-clusters that includes 
MACS J1226.8+2153 which is centered $\simeq 2.5'$ to 
the south. A summary of 
the source properties is given in Table 1.

\subsection{Image and Photometric Calibration}

In preparation for initial spectroscopy of probable arcs, 
deeper multi-band imaging was acquired in addition to 
pre-existing SGAS discovery images. Pre-imaging exposures 
totaling 5 minutes in length in $g$, $r$, $i$ acquired 
with the GMOS instrument on the 8m Gemini North telescope 
were used to design slit-masks for the cluster SDSS J1226+2152. 
Images were reduced using a customized pipeline described 
elsewhere. A composite color image is shown in Figure 1. 
SDSS J1527+0652 serendipitously falls in a Red-Sequence 
Cluster Survey-2 (RCS2) field, which supplies 4, 8 and 6 
min exposures in $g$, $r$, $z$ with MegaCam on the 3.6m 
CFHT telescope; the composite color image is also shown in Figure 1.

The images are transformed to a common reference frame, the 
$i$-band for SGAS 122651+215220 and the $r$-band for SGAS J152745.1+065219, 
and calibrated against the Sloan Digital Sky Survey. We construct 
an empirical, normalized point spread function (PSF) for each 
image based on a well-defined, non-saturated, isolated reference 
star. Since the three optical bands have comparable seeing, we can 
safely ignore PSF mismatches between them. Apertures 
are created by first defining a curve along the long axis of 
the extended source and convolving this curve with the appropriate 
PSF. The effective aperture is then defined 
as a contour of this convolution that is at 
exactly 2.5 times the FWHM of the PSF.

Both sources lie close to one or more members of the foreground 
galaxy cluster. We use the GALFIT package (Peng et al. 2002) to 
fit a Sersic profile to these cluster members in each of the images 
and subtract their flux. 
 
After carefully removing the sky level and masking any remaining 
outliers in the outskirts of the apertures (pixel values more 
extreme than $\pm 5 \sigma$), the final magnitude is measured 
in the effective aperture defined above and corrected to 
an equivalent radius of 6" based on the curve of growth of the PSF star.

\begin{figure}
\centering
\includegraphics[scale=0.5]{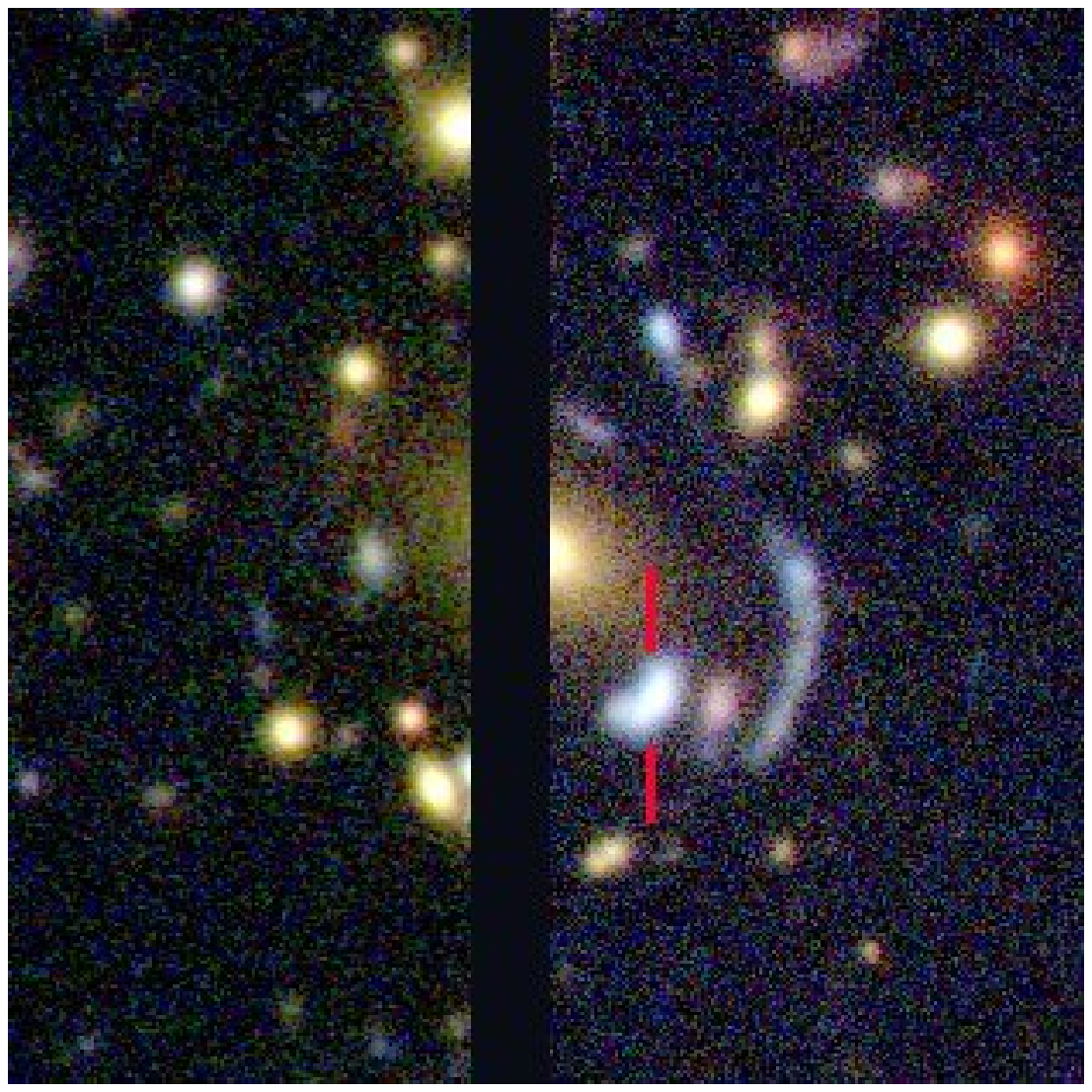}
\includegraphics[scale=0.665]{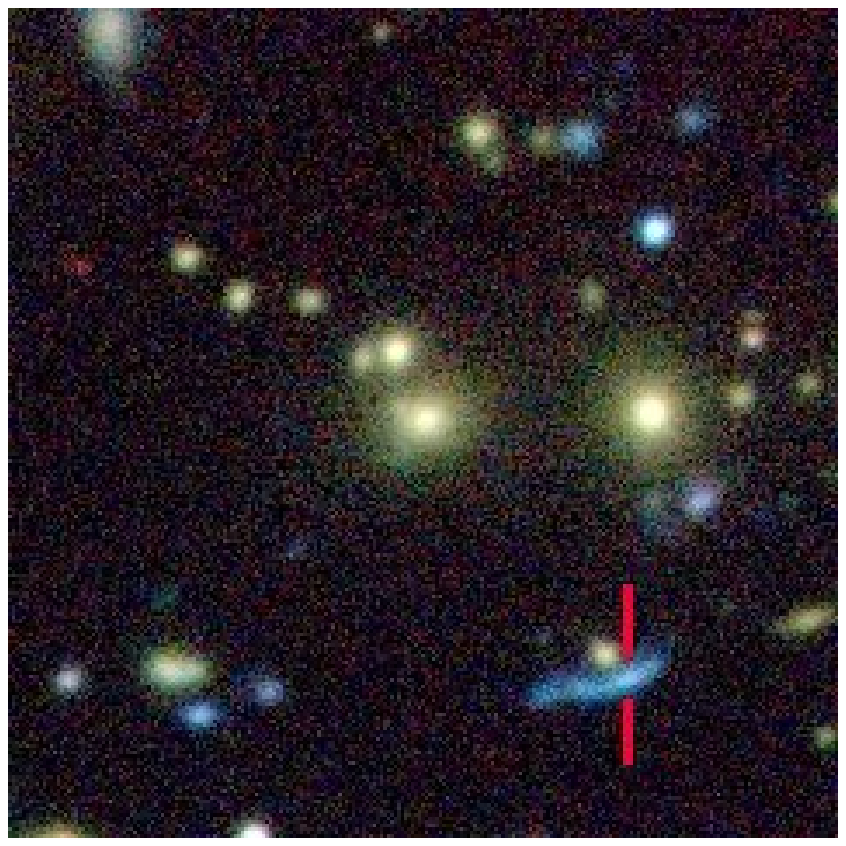}
\caption{\emph{Top panel}: SGAS J122651.3+215220 at $z=2.9233$ 
from Gemini GMOS in $g$,$r$,$i$ (45''/side). Part of the BCG falls
on a GMOS chip gap. The lensed LBG is highlighted by the
red cross.
\emph{Bottom Panel}: SGAS J152745.1+065219 at $z=2.7593$ from MegaCam in $g$,$r$,$z$,
(45''/side).}
\end{figure}
\subsection{Spectroscopy}

Spectra (Figure 2) of the 3100-8000~\AA\ region were obtained with the Mage spectrograph 
\citep{marshall08} on the Magellan Clay 6.5m telescope, using a 2~\arcsec\ slit.  
The spectrum for SGAS J122651.3+215220  is a combination of three observations from 09 Feb 2008, 22 Apr 2009, 
23 Apr 2009 UT, with a total integration of 6.1 hr.  The spectrum for SGAS J152745.1+065219 
is from the night of 23 Apr 2009 UT, with a total integration of 1.5 hr.  The data were bias-subtracted, flat-fielded, 
sky-subtracted, optimally extracted, and wavelength-corrected following \citet{kelson03}. 
Each slit was aligned with the 
parallactic angle, and positioned at the peak brightness in g-band.

Each spectrum shows damped Ly$\alpha$ along with
characteristic UV metal absorption lines 
(C{\sc II}, Si{\sc II}, Si{\sc III}, Si{\sc IV}) all consistent with redshifts of 
$z=2.7593$ and $z=2.9233$, for SGAS J152745.1+065219 and SGAS J122651.3+215220, respectively.

\begin{figure*}[t]
\centering
\includegraphics[scale=0.60]{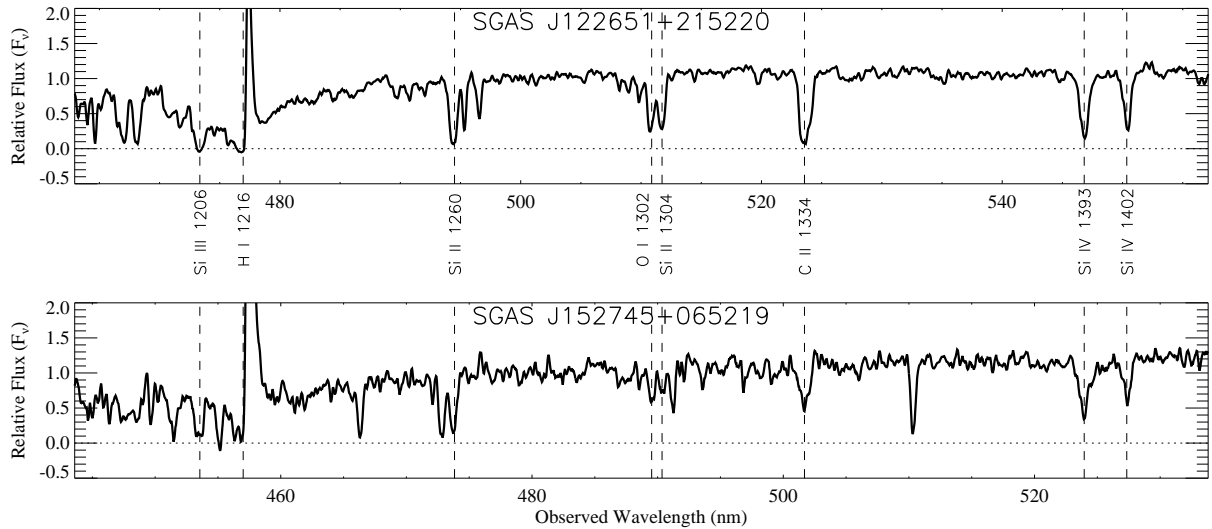}
\caption{$R=3000$ MagE spectra of 2 new lensed-LBGs. The spectra are smoothed for presentation. 
\emph{Top panel}: SGAS J122651.3+215220 at $z=2.9233$ . 
\emph{Bottom Panel}: SGAS J152745.1+065219 at $z=2.7593$.}
\end{figure*}

\section{Lens Properties and Mass Modeling}

To understand luminosity-dependent source properties, a lens mass model 
is needed to correct for the magnification effect, $M$.
The redshift, morphology, and position of the lensed LBG relative to the cluster center
permit simple modeling of the cluster lens, as has been done in other systems. 
However, unlike the Cosmic Eye and the 8 o'clock arc, neither LBG is multiply-imaged, 
which reduces the robustness of the lens model. Moreover, the fact that 
these lenses were originally selected as massive galaxy
clusters highlights the complexity of the lens relative to lensing by a single galaxy. 
The lens modeling is executed via $\tt{lenstool}$ \citep{jullo07}.

The cluster SDSS J1226+2152 (Figure 3) is dominated by a central BCG, 
around which several lensed features are detected. As part of our spectroscopic
follow-up campaign, redshifts were secured for some of these background sources. 
In addition to the LBG, SGAS J122651.3+215220, at $z=2.9233$, we have 
identified a giant arc at $z=2.923$, a source that \citet{wen09} noted is 
a ``probable'' lensed galaxy. Other background sources appear at $z=0.772$ and $z=0.732$. 
Possible counter-images for any of these sources, if they exist, were not identified in the data.

The SDSS J1226+1252 lens model is composed of a cluster halo that is represented by an NFW \citep{nfw96} profile; 
the BCG is represented by a pseudo-isothermal ellipsoid mass distribution \citep[PIEMD;][]{jullo07}, 
with parameters that follow the observed light distribution of the galaxy; and 
external shear is also included from the neighboring cluster MACS J1226.8+2153, $153.12''$ south and $8.6''$ 
west of the BCG, represented as a circular PIEMD with $\sigma_v\sim1000$ km s$^{-1}$.
We allow all the parameters of the NFW halo to vary, as well as the velocity dispersion of the BCG and the external shear. 
The best-fit model is determined using MCMC, and the minimization is done in the source plane.
We use the redshift and position of the giant arc and its counter image as constraints. We also force 
a critical curve at the 
location of the LBG, assuming  that the observed arc is a result of a merging pair. These models typically 
predict a magnification of $M\gtrsim 40$. 

\begin{figure}
\centering
\includegraphics[scale=0.30]{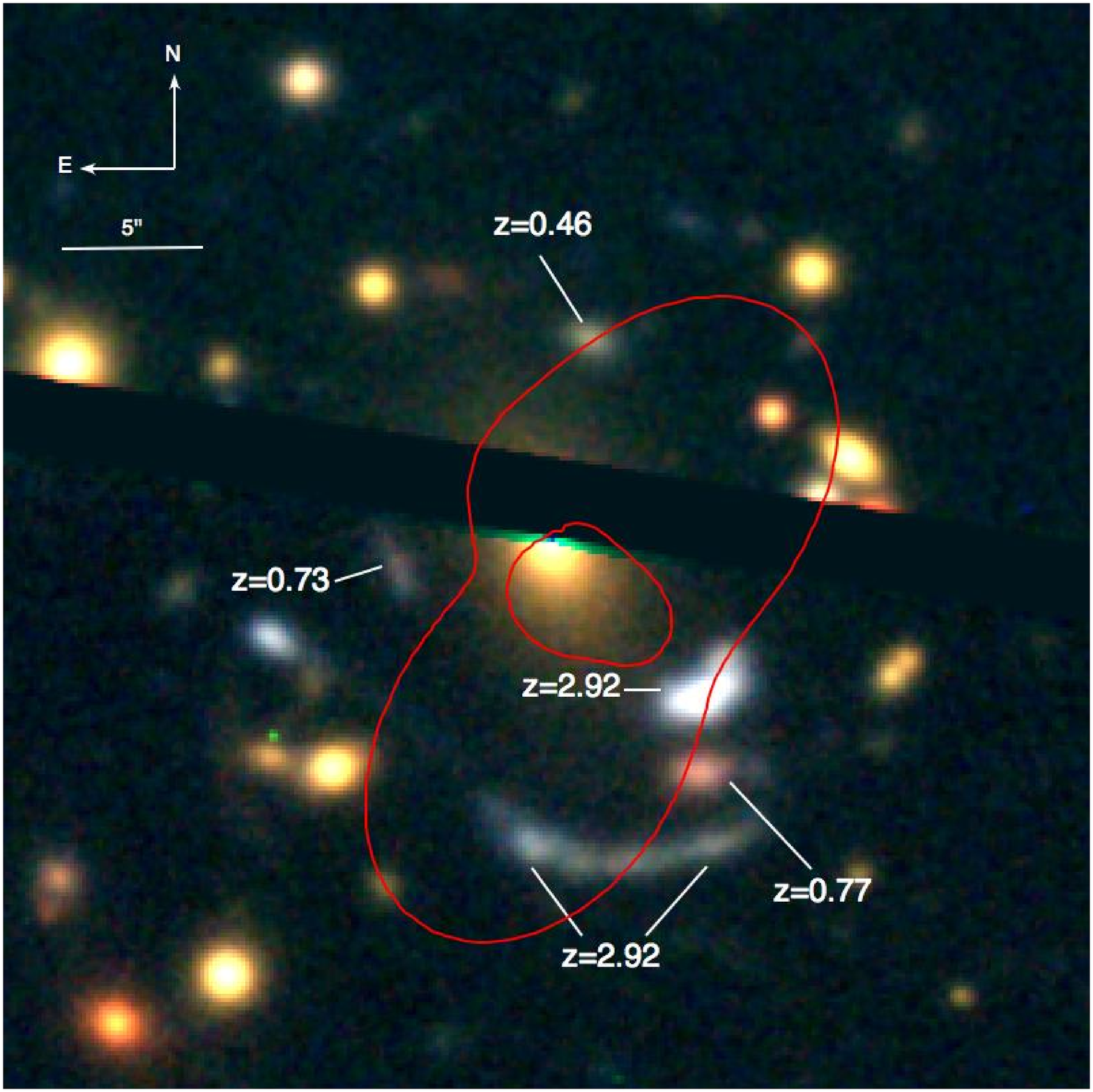}
\caption{SDSS J1226+2152 Lens Model. 
Background objects with redshifts are labeled, including
the newly discovered LBG SGAS J122651.3+215220; to the south
a source at the same redshift appears lensed as a giant arc.
Source redshifts and positions are used to construct 
a PIEMD mass model (see text), whose radial 
and tangential critical curves (inner and outer red lines)
are overplotted. } 
\end{figure}

In the case of  SDSS J1527+2152 there are fewer lensing constraints, as no other lensed 
features are accessible in RCS-2 imaging.
Furthermore, the core of SDSS J1527+2152 is dominated by two bright galaxies of approximately 
the same brightness -- not by a dominant BCG -- preventing the common assumption 
that the cluster halo is centered on the BCG.
We therefore do not attempt to fit a model to the cluster, but instead we explore a large set of lens 
models spanning a wide range in parameter values.
We take advantage of the known cluster velocity dispersion, $\sigma_v \simeq 908$ km s$^{-1}$, 
and represent the cluster with a PIEMD. The parameter space we explore includes the position of the halo, 
its ellipticity and PA, and the core and cut radii. We examine two families of models. In the 
simplest case, we assume that the halo is circular, and centered on a line connecting the two central galaxies.
For each model we compute the locations and relative magnifications of predicted counter images for the LBG.
Since we do not observe a counter images at RCS-2 depths, we rule out models that predict such counter images. 
We find that in the allowed models of this family, the magnification is constrained to be under $\sim15$.
However, the more complicated case, which allows models with high ellipticity, the requirement 
that the model does not produce counter images for the LBG does not constrain the magnification.
Hence, we proceed by taking $M=15$ as an upper bound on the magnification of SGAS J152745.1+065219.

\begin{deluxetable*}{lll}
\tablecaption{Lens Properties, Arc Photometry.\label{tab}}
\tablewidth{0pt}
\tabletypesize{\tiny}
\tablehead{
\colhead{} & 
\colhead{SGAS J122651.3+215220\phantom{fillllllllll}} & 
\colhead{SGAS J152745.1+065219\phantom{fillllllllll}}
}
\startdata
Cluster Lens & SDSS J1226+2152 & SDSS J1527+0652\\
Cluster Centroid \tablenotemark{1}($\alpha,\delta$) & (12:26:51.74,+21:52:25.57) & (15:27:45.15,+06:52:25.70)\\
Arc ($\alpha,\delta$) &      (12:26:51.34,+21:52:20.26) & (15:27:45.17,+06:52:19.17)\\
Arc Total Mag\tablenotemark{2} (g,r,i/z) & (21.14,20.60,20.51) & (20.90,20.52,20.58)\\
Lens Magnification      & $\simeq 40$   & $\lesssim 15$ \\
Arc Source Mag\tablenotemark{3} (g,r,i/z) & (25.16,24.61,24.52) & (23.90,23.52,23.5)\\
Arc $\mu$ (mag arcsec$^{-2}$) & (23.78,23.11,22.81)  & (25.15,24.52,24.12)\\

\enddata
\tablenotetext{1}{The BCG coordinates are given for SDSS J1226+2152, while the luminosity-weighted centroid coordinates are used for SDSS J1527+0652.}
\tablenotetext{2}{SGAS J122651.3+215220 is measured in $g,r,i$ and SGAS J152745.1+065219 in $g,r,z$; errors on magnitudes are typically $\simeq 0.05$ mags.}
\tablenotetext{3}{Source magnitude corrected for assumed lens magnification. }
\end{deluxetable*}

\section{Source Properties}
To place these and other lensed-LBGs in context, we compare to the LBG sample presented 
in \citet{steidel03} (S03).
That catalog is constructed from a photometric survey that 
targeted $z \sim 3$ LBGs. The comparison sample includes a subset of 237 objects 
that pass the photometric selection criteria, are classified as galaxies, 
and have emission and absorption line redshifts that agree to within 0.02 in redshift. 
Absent the lensing interpretation, both LBGs in this study sit 2 magnitudes
brighter, or about a factor of $\sim 6$ brighter than the brightest object 
in that sample.  

The apparent arc magnitudes in each band are given in Table 1, as are the 
lens magnifications, the lensing-corrected apparent magnitudes, and coordinates.
Because the $r$-band samples the continuum redward of both the characteristic 
strong Ly-$\alpha$ emission and Lyman break spectral features \citep[e.g.,][]{steidel95}, 
we consider only $r$-band quantities in all comparisons.
Using the published magnifications and observed total magnitudes of the  
the brightest of the multiple images of both the 8 o'clock arc and the Cosmic Eye, 
the lensing-corrected apparent magnitudes
are $r=22.93$ and $r=24.16$, respectively. Assuming a magnification, $M=15$, 
SGAS J152745.1+065219 has a similar
lensing-corrected $r$-band magnitude of $r=23.52$. However, SGAS J122651.3+215220,  
$r=24.61$, is about a magnitude fainter. Thus, while all four sources are at 
similar redshifts, it is evident that SGAS J122651.3+215220 is quite faint, and 
is in fact similar in brightness to cB58, which experiences 
a magnification of $M\simeq30$ \citep{seitz98,pettini00}, or a lensing-corrected magnitude of $r=24.77$.

When placed in the context of the S03 LBG survey, it is apparent that the 
recently discovered sources include objects that form a sparse, but wide-ranging
sample of the underlying LBG population. In the brightest quartile, the source galaxy 
of the 8 o'clock arc is among the $\simeq 10\%$ brightest LBGs in S03 in both 
magnitude and luminosity, and SGAS J152745.1+065219 is among the brightest
25\% . The Cosmic Eye and cB58 both are found in the third quartile, 
and SGAS J122651.3+215220, the dimmest of these lensed-LBGs, is fainter than
$80\%$ of the galaxies in the S03 sample.

\section{Conclusion}
While we require additional deep imaging to better
constrain the lens models for SGAS J122651.3+215220 and SGAS J152745.1+065219, the existing MagE spectra, other 
spectroscopy, and additional IR and UV imaging form a basis for a series of forthcoming papers 
that investigate the intrinsic properties of the LBGs in this study (Rigby et al., 2010).

These discoveries nearly double the number of known lensed LBGs. The ambitious follow-up program
built into SGAS will continue to grow the $z \sim 3$ lensed-LBG sample as well as
other lensed source populations (e.g., Ly-$\alpha$ emitters, Bayliss et al., 2010).
The systematic nature of this program offers the possibility
of building large samples of strongly-lensed high redshift galaxies whose high fluxes
enable both the acquisition of high S/N spectra and a look at the faint end of the luminosity function.

\acknowledgements Based on observations obtained at the Gemini Observatory, which is operated by the
Association of Universities for Research in Astronomy, Inc., under a cooperative agreement
with the NSF on behalf of the Gemini partnership: the National Science Foundation (United
States), the Science and Technology Facilities Council (United Kingdom), the
National Research Council (Canada), CONICYT (Chile), the Australian Research Council
(Australia), Ministério da Ciência e Tecnologia (Brazil) 
and Ministerio de Ciencia, Tecnología e Innovación Productiva  (Argentina)

Based on observations obtained with MegaPrime/MegaCam, a joint project of CFHT and CEA/DAPNIA, 
at the Canada-France-Hawaii Telescope (CFHT) which is operated by the National Research Council 
(NRC) of Canada, the Institut National des Science de l'Univers of the Centre National de la Recherche 
Scientifique (CNRS) of France, and the University of Hawaii. This work is based in part on data 
products produced at TERAPIX and the Canadian Astronomy Data Centre as part of the 
Canada-France-Hawaii Telescope Legacy Survey, a collaborative project of NRC and CNRS. 

Based on observations made with the Nordic Optical Telescope, operated
on the island of La Palma jointly by Denmark, Finland, Iceland,
Norway, and Sweden, in the Spanish Observatorio del Roque de los
Muchachos of the Instituto de Astrofisica de Canarias.

\bibliographystyle{apj}
\bibliography{brightarcs}

\begin{thebibliography}{29}
\expandafter\ifx\csname natexlab\endcsname\relax\def\natexlab#1{#1}\fi

\bibitem[{{Adelberger} {et~al.}(1998){Adelberger}, {Steidel}, {Giavalisco},
  {Dickinson}, {Pettini}, \& {Kellogg}}]{adelberger98}
{Adelberger}, K.~L., {Steidel}, C.~C., {Giavalisco}, M., {Dickinson}, M.,
  {Pettini}, M., \& {Kellogg}, M. 1998, \apj, 505, 18

\bibitem[{{Adelberger} {et~al.}(2003){Adelberger}, {Steidel}, {Shapley}, \&
  {Pettini}}]{adelberger03}
{Adelberger}, K.~L., {Steidel}, C.~C., {Shapley}, A.~E., \& {Pettini}, M. 2003,
  \apj, 584, 45

\bibitem[{{Allam} {et~al.}(2007){Allam}, {Tucker}, {Lin}, {Diehl}, {Annis},
  {Buckley-Geer}, \& {Frieman}}]{allam07}
{Allam}, S.~S., {Tucker}, D.~L., {Lin}, H., {Diehl}, H.~T., {Annis}, J.,
  {Buckley-Geer}, E.~J., \& {Frieman}, J.~A. 2007, \apjl, 662, L51

\bibitem[{{Cabanac} {et~al.}(2007){Cabanac}, {Alard}, {Dantel-Fort}, {Fort},
  {Gavazzi}, {Gomez}, {Kneib}, {Le F{\`e}vre}, {Mellier}, {Pello}, {Soucail},
  {Sygnet}, \& {Valls-Gabaud}}]{cabanac07}
{Cabanac}, R.~A. {et~al.} 2007, \aap, 461, 813

\bibitem[{{Coppin} {et~al.}(2007){Coppin}, {Swinbank}, {Neri}, {Cox}, {Smail},
  {Ellis}, {Geach}, {Siana}, {Teplitz}, {Dye}, {Kneib}, {Edge}, \&
  {Richard}}]{coppin07}
{Coppin}, K.~E.~K. {et~al.} 2007, \apj, 665, 936

\bibitem[{{Finkelstein} {et~al.}(2009){Finkelstein}, {Papovich}, {Rudnick},
  {Egami}, {LeFloc'h}, {Rieke}, {Rigby}, \& {Willmer}}]{finkelstein09}
{Finkelstein}, S.~L., {Papovich}, C., {Rudnick}, G., {Egami}, E., {LeFloc'h},
  E., {Rieke}, M.~J., {Rigby}, J.~R., \& {Willmer}, C.~N.~A. 2009, \apj, 700,
  376

\bibitem[{{Gladders}(2005)}]{gladders05}
{Gladders}, M.~D. 2005, in IAU Symposium, Vol. 225, Gravitational Lensing
  Impact on Cosmology, ed. {Y.~Mellier \& G.~Meylan}, 149--154

\bibitem[{{Gladders} \& {Yee}(2000)}]{gladders00}
{Gladders}, M.~D., \& {Yee}, H.~K.~C. 2000, \aj, 120, 2148

\bibitem[{{Hennawi} {et~al.}(2008){Hennawi}, {Gladders}, {Oguri}, {Dalal},
  {Koester}, {Natarajan}, {Strauss}, {Inada}, {Kayo}, {Lin}, {Lampeitl},
  {Annis}, {Bahcall}, \& {Schneider}}]{hennawi08}
{Hennawi}, J.~F. {et~al.} 2008, \aj, 135, 664

\bibitem[{{Jullo} {et~al.}(2007){Jullo}, {Kneib}, {Limousin},
  {El{\'{\i}}asd{\'o}ttir}, {Marshall}, \& {Verdugo}}]{jullo07}
{Jullo}, E., {Kneib}, J., {Limousin}, M., {El{\'{\i}}asd{\'o}ttir}, {\'A}.,
  {Marshall}, P.~J., \& {Verdugo}, T. 2007, New Journal of Physics, 9, 447

\bibitem[{{Kelson}(2003)}]{kelson03}
{Kelson}, D.~D. 2003, \pasp, 115, 688

\bibitem[{{Marshall} {et~al.}(2008){Marshall}, {Burles}, {Thompson},
  {Shectman}, {Bigelow}, {Burley}, {Birk}, {Estrada}, {Jones}, {Smith},
  {Kowal}, {Castillo}, {Storts}, \& {Ortiz}}]{marshall08}
{Marshall}, J.~L. {et~al.} 2008, in Presented at the Society of Photo-Optical
  Instrumentation Engineers (SPIE) Conference, Vol. 7014, Society of
  Photo-Optical Instrumentation Engineers (SPIE) Conference Series

\bibitem[{{Mehlert} {et~al.}(2001){Mehlert}, {Seitz}, {Saglia}, {Appenzeller},
  {Bender}, {Fricke}, {Hoffmann}, {Hopp}, {Kudritzki}, \&
  {Pauldrach}}]{mehlert01}
{Mehlert}, D. {et~al.} 2001, \aap, 379, 96

\bibitem[{{Navarro} {et~al.}(1996){Navarro}, {Frenk}, \& {White}}]{nfw96}
{Navarro}, J.~F., {Frenk}, C.~S., \& {White}, S.~D.~M. 1996, \apj, 462, 563

\bibitem[{{Nesvadba} {et~al.}(2006){Nesvadba}, {Lehnert}, {Eisenhauer},
  {Genzel}, {Seitz}, {Davies}, {Saglia}, {Lutz}, {Tacconi}, {Bender}, \&
  {Abuter}}]{nesvadba06}
{Nesvadba}, N.~P.~H. {et~al.} 2006, \apj, 650, 661

\bibitem[{{Pettini} {et~al.}(2002){Pettini}, {Rix}, {Steidel}, {Adelberger},
  {Hunt}, \& {Shapley}}]{pettini02}
{Pettini}, M., {Rix}, S.~A., {Steidel}, C.~C., {Adelberger}, K.~L., {Hunt},
  M.~P., \& {Shapley}, A.~E. 2002, \apj, 569, 742

\bibitem[{{Pettini} {et~al.}(2000){Pettini}, {Steidel}, {Adelberger},
  {Dickinson}, \& {Giavalisco}}]{pettini00}
{Pettini}, M., {Steidel}, C.~C., {Adelberger}, K.~L., {Dickinson}, M., \&
  {Giavalisco}, M. 2000, \apj, 528, 96

\bibitem[{{Seitz} {et~al.}(1998){Seitz}, {Saglia}, {Bender}, {Hopp}, {Belloni},
  \& {Ziegler}}]{seitz98}
{Seitz}, S., {Saglia}, R.~P., {Bender}, R., {Hopp}, U., {Belloni}, P., \&
  {Ziegler}, B. 1998, \mnras, 298, 945

\bibitem[{{Shapley} {et~al.}(2003){Shapley}, {Steidel}, {Pettini}, \&
  {Adelberger}}]{shapley03}
{Shapley}, A.~E., {Steidel}, C.~C., {Pettini}, M., \& {Adelberger}, K.~L. 2003,
  \apj, 588, 65

\bibitem[{{Siana} {et~al.}(2009){Siana}, {Smail}, {Swinbank}, {Richard},
  {Teplitz}, {Coppin}, {Ellis}, {Stark}, {Kneib}, \& {Edge}}]{siana09}
{Siana}, B. {et~al.} 2009, \apj, 698, 1273

\bibitem[{{Siana} {et~al.}(2008){Siana}, {Teplitz}, {Chary}, {Colbert}, \&
  {Frayer}}]{siana08}
{Siana}, B., {Teplitz}, H.~I., {Chary}, R.-R., {Colbert}, J., \& {Frayer},
  D.~T. 2008, \apj, 689, 59

\bibitem[{{Smail} {et~al.}(2007){Smail}, {Swinbank}, {Richard}, {Ebeling},
  {Kneib}, {Edge}, {Stark}, {Ellis}, {Dye}, {Smith}, \& {Mullis}}]{smail07}
{Smail}, I. {et~al.} 2007, \apjl, 654, L33

\bibitem[{{Steidel} {et~al.}(2003){Steidel}, {Adelberger}, {Shapley},
  {Pettini}, {Dickinson}, \& {Giavalisco}}]{steidel03}
{Steidel}, C.~C., {Adelberger}, K.~L., {Shapley}, A.~E., {Pettini}, M.,
  {Dickinson}, M., \& {Giavalisco}, M. 2003, \apj, 592, 728

\bibitem[{{Steidel} {et~al.}(1996){Steidel}, {Giavalisco}, {Pettini},
  {Dickinson}, \& {Adelberger}}]{steidel96a}
{Steidel}, C.~C., {Giavalisco}, M., {Pettini}, M., {Dickinson}, M., \&
  {Adelberger}, K.~L. 1996, \apjl, 462, L17+

\bibitem[{{Steidel} {et~al.}(1995){Steidel}, {Pettini}, \&
  {Hamilton}}]{steidel95}
{Steidel}, C.~C., {Pettini}, M., \& {Hamilton}, D. 1995, \aj, 110, 2519

\bibitem[{{Teplitz} {et~al.}(2000){Teplitz}, {McLean}, {Becklin}, {Figer},
  {Gilbert}, {Graham}, {Larkin}, {Levenson}, \& {Wilcox}}]{teplitz00}
{Teplitz}, H.~I. {et~al.} 2000, \apjl, 533, L65

\bibitem[{{Wen} {et~al.}(2009){Wen}, {Han}, {Xu}, {Jiang}, {Guo}, {Wang}, \&
  {Liu}}]{wen09}
{Wen}, Z., {Han}, J., {Xu}, X., {Jiang}, Y., {Guo}, Z., {Wang}, P., \& {Liu},
  F. 2009, Research in Astronomy and Astrophysics, 9, 5

\bibitem[{{Yee} {et~al.}(1996){Yee}, {Ellingson}, {Bechtold}, {Carlberg}, \&
  {Cuillandre}}]{yee96}
{Yee}, H.~K.~C., {Ellingson}, E., {Bechtold}, J., {Carlberg}, R.~G., \&
  {Cuillandre}, J. 1996, \aj, 111, 1783

\bibitem[{{York} {et~al.}(2000){York}, {Adelman}, {Anderson}, {Anderson},
  {Annis}, {Bahcall}, {Bakken}, {Barkhouser}, {Bastian}, {Berman}, {Boroski},
  {Bracker}, {Briegel}, {Briggs}, {Brinkmann}, {Brunner}, {Burles}, {Carey},
  {Carr}, {Castander}, {Chen}, {Colestock}, {Connolly}, {Crocker}, {Csabai},
  {Czarapata}, {Davis}, {Doi}, {Dombeck}, {Eisenstein}, {Ellman}, {Elms},
  {Evans}, {Fan}, {Federwitz}, {Fiscelli}, {Friedman}, {Frieman}, {Fukugita},
  {Gillespie}, {Gunn}, {Gurbani}, {de Haas}, {Haldeman}, {Harris}, {Hayes},
  {Heckman}, {Hennessy}, {Hindsley}, {Holm}, {Holmgren}, {Huang}, {Hull},
  {Husby}, {Ichikawa}, {Ichikawa}, {Ivezi{\'c}}, {Kent}, {Kim}, {Kinney},
  {Klaene}, {Kleinman}, {Kleinman}, {Knapp}, {Korienek}, {Kron}, {Kunszt},
  {Lamb}, {Lee}, {Leger}, {Limmongkol}, {Lindenmeyer}, {Long}, {Loomis},
  {Loveday}, {Lucinio}, {Lupton}, {MacKinnon}, {Mannery}, {Mantsch}, {Margon},
  {McGehee}, {McKay}, {Meiksin}, {Merelli}, {Monet}, {Munn}, {Narayanan},
  {Nash}, {Neilsen}, {Neswold}, {Newberg}, {Nichol}, {Nicinski}, {Nonino},
  {Okada}, {Okamura}, {Ostriker}, {Owen}, {Pauls}, {Peoples}, {Peterson},
  {Petravick}, {Pier}, {Pope}, {Pordes}, {Prosapio}, {Rechenmacher}, {Quinn},
  {Richards}, {Richmond}, {Rivetta}, {Rockosi}, {Ruthmansdorfer}, {Sandford},
  {Schlegel}, {Schneider}, {Sekiguchi}, {Sergey}, {Shimasaku}, {Siegmund},
  {Smee}, {Smith}, {Snedden}, {Stone}, {Stoughton}, {Strauss}, {Stubbs},
  {SubbaRao}, {Szalay}, {Szapudi}, {Szokoly}, {Thakar}, {Tremonti}, {Tucker},
  {Uomoto}, {Vanden Berk}, {Vogeley}, {Waddell}, {Wang}, {Watanabe},
  {Weinberg}, {Yanny}, \& {Yasuda}}]{york00}
{York}, D.~G. {et~al.} 2000, \aj, 120, 1579

\end{thebibliography}

\end{document}